\documentclass[journal]{IEEEtran}
\usepackage[utf8]{inputenc}
\usepackage{amsmath}
\usepackage{graphicx}
\usepackage[mathscr]{euscript}
\usepackage{bm}
\usepackage{cite}
\usepackage{xcolor}
\usepackage[mathscr]{euscript}
\usepackage{csquotes}
\usepackage{url}

\title{Generalized Helmholtz Decomposition for Modal Analysis of Electromagnetic Problems in Inhomogeneous Media}
\author{
Jie Zhu,
Thomas E. Roth,~\IEEEmembership{Member,~IEEE,}\\
Dong-Yeop Na,~\IEEEmembership{Member,~IEEE,}
and 
Weng Cho Chew,~\IEEEmembership{Life Fellow,~IEEE}
}

\renewcommand{\epsilon}{\varepsilon}

\begin{document}

\maketitle

\begin{abstract}
Potential-based formulation with generalized Lorenz gauge can be used in the quantization of electromagnetic fields in inhomogeneous media \cite{na2020quantum}.
However, one often faces the redundancy of modes when finding eigenmodes from potential-based formulation.
In free space, this can be explained by the connection to the well-known Helmholtz decomposition.
In this work, we generalize the Helmholtz decomposition to its generalized form, echoing the use of generalized Lorenz gauge in inhomogeneous media \cite{chew2014vector}. 
We formulate electromagnetics eigenvalue problems using vector potential formulation which is often used in numerical quantization \cite{na2020quantum}. 
The properties of the differential operators are mathematically analyzed. 
Orthogonality relations between the two classes of modes are proved in both continuous and discrete space. 
Completeness of two sets of modes and the orthogonality relations are numerically validated in inhomogeneous anisotropic media. 
This work serves as a foundation for numerical quantization of electromagnetic fields in inhomogeneous media with potential-based formulation. 
\end{abstract}

\section{Introduction}
The formulation of electromagnetic theory based on \textbf{E}, \textbf{H}, \textbf{D}, and \textbf{B} offers physical insight that has resulted in the development of many electromagnetic-related technologies. 
However, classical electromagnetic equations formulated in terms of \textbf{E}-\textbf{H} have low-frequency or long-wavelength breakdown \cite{chew2014vector}. Many numerical methods based on \textbf{E}-\textbf{H} formulation are unstable at low frequencies. Therefore, \textbf{E}-\textbf{H} formulation is not truly multi-scale, but exhibits catastrophic breakdown when the dimensions of objects become much smaller than the local wavelength, or when the frequency is low. 

Much previous work has been devoted to eliminating low-frequency breakdown of \textbf{E}-\textbf{H} formulation. 
For differential equation solvers, Manges and Cendes \cite{manges1995generalized} proposed a generalized tree-cotree gauge to eliminate the matrix null space of the magnetostatic curl-curl equation.
In another work \cite{lee2003hierarchical}, hierarchical basis functions and tree-cotree splitting are used to enhance the stability of the finite-element procedure.
For time domain solvers, method based on changing curl-curl operator to Laplacian is proposed to handle the ill-conditioning of the system matrix \cite{xue2021fast}.
For integral equation solvers, loop-tree decomposition was first proposed which separates the electrostatic and magnetostatic physics \cite{wilton1993novel,burton1995study,wu1995study,vecchi1999loop,zhao2000integral,mautz1984field}. Augmented electric field integral equation was then proposed which eliminated the need to search for loops \cite{qian2008augmented}.

Electromagnetic formulation based on potential theory has received increasing attention recently since it eliminates the low-frequency breakdown issue elegantly. 
It removes the null space solution by converting singular matrices to full-rank Laplacian-like matrices.
In a homogeneous medium, it is easy to find its connection to the Helmholtz decomposition.
A number of previous works have dealt with potential-based (or $\mathbf{A}$-$\Phi$) formulation with both differential equations \cite{li2016finite,dyczij1998fast,ryu2016finite,roth2022theory}, and integral equations \cite{de2003volume,vico2016decoupled,liu2018potential,li2018decoupled,khalichi2019broadband,roth2018development,roth2021lorenz,chen2022low}. 
Moreover, potential-based formulation is more compatible with quantum theory. Aharonov-Bohm effect is an example in quantum theory where \textbf{E}-\textbf{H} are zero, but vector potential \textbf{A} is not zero, and yet, the effect of \textbf{A} is felt \cite{aharonov1959significance}. 
Besides, in many quantum optics studies, both vector potential $\mathbf{A}$ and scalar potential $\Phi$ are used when the electromagnetic fields are incorporated into the Schr\"{o}dinger equation \cite{ryu2016finite}. 

To perform numerical quantization of electromagnetic fields in the mode picture with potential-based formulation, it is often needed to find the electromagnetic eigenmodes of the system with inhomogeneous media by numerical methods \cite{na2020quantum,na2021diagonalization,chew2016quantum2}.
When solving potential-based formulation as electromagnetics eigenvalue problems, null space solution does not exist, but one often obtains some ``extra solutions" that feature zero $\mathbf{E}$ and $\mathbf{H}$ even though $\mathbf{A}$ and $\Phi$ are nonzero. 
In a homogeneous medium, this can be explained by making connection to the Helmholtz decomposition, as will be shown in Section \ref{section:motivation}. 

Due to the need of numerical quantization in complex media in modern quantum technology \cite{na2020quantum,na2021diagonalization}, an understanding of the solution space of vector potential wave equation in general inhomogeneous media is called for. 
The connection between the extra modes and the solution space should be made. 
A general scheme is needed to explain it in inhomogeneous media.

In this work, we investigate how to generalize Helmholtz decomposition to inhomogeneous media. We formulate electromagnetics eigenvalue problems using the vector potential formulation, and investigate the modes from solving the system equations. The formulation can be readily used in the numerical quantization scheme in quantum electromagnetics \cite{na2020quantum,na2021diagonalization,chew2016quantum2}. 
We first review Helmholtz theorem and motivate the need for its generalization. We then begin the proof by analyzing the differential operator associated with the wave equation for vector potential in Section \ref{section:motivation}. 
In Section \ref{section:orthogonality}, we define div-$\epsilon$-free modes in inhomogeneous media, which is the counterpart of div-free modes in homogeneous case. 
We prove analytically that the div-$\epsilon$-free modes are orthogonal to the curl-free modes in source-less inhomogeneous regions, by both continuous-space and discrete-space calculus. No such proof has been found in the literature. The proof can be readily generalized to anisotropic media.
We provide numerical results in Section \ref{section:numerical}: We first show numerically that Helmholtz decomposition can be extended to inhomogeneous case, which we call generalized Helmholtz decomposition (GHD). We then show numerically the orthogonality condition. We conclude the work in Section \ref{section:conclusion}.

\section{A Review of Helmholtz Theorem and Motivation of Generalization}\label{section:motivation}
\subsection{A Traditional Proof}
Two important theorems establish the conditions for the existence and uniqueness of solutions to time-independent problems in electromagnetic theory. The first one is \cite{arfken2013mathematical}: 
\begin{displayquote}
\textit{A vector field is uniquely specified by giving its divergence and its curl within a simply connected region and its normal component on the boundary}.
\end{displayquote}
It can be proved by contradiction. Assume that two different vector fields $\mathbf{A}$ and $\mathbf{A'}$ have the same divergence and curl
\begin{align}
    \nabla\cdot\mathbf{A} &= \nabla\cdot\mathbf{A'} = s  \\
    \nabla\times\mathbf{A} &= \nabla\times\mathbf{A'} = \mathbf{c} 
\end{align}
and their normal component $\mathbf{A}_n$ and $\mathbf{A}'_n$ on the boundary are also the same. Let $\mathbf{F}=\mathbf{A}-\mathbf{A'}$. Then $\nabla\cdot\mathbf{F}$, $\nabla\times\mathbf{F}$, and $\mathbf{F}_n$ are all zero. Thus there exists a $\phi$ such that $\mathbf{F}=-\nabla\phi$ and it satisfies
\begin{align}
    \nabla^2\phi=0 . \label{laplace eq}
\end{align}
Due to Earnshaw's theorem, the solution to (\ref{laplace eq}) must have its maximum and minimum on the boundary.
Together with the condition $\mathbf{F}_n=0$, or $\dfrac{\partial \phi}{\partial n}=0$, it must have that $\phi$ is a constant everywhere. Hence $\mathbf{F}=0$ everywhere. Thus $\mathbf{A}$ and $\mathbf{A'}$ are identical.  

Given the proof of uniqueness, we then review the Helmholtz theorem. The Helmholtz theorem states \cite{arfken2013mathematical,collin2007foundations,jackson1999classical}: 
\begin{displayquote}
\textit{A vector $\mathbf{A}$ with both source and circulation densities vanishing at infinity may be written as the sum of two parts, one of which is irrotational, curl-free, or longitudinal, the other of which is solenoidal,  div-free, or transverse. }
\end{displayquote}
That is,
\begin{align}
    \mathbf{A} = \mathbf{A}_\parallel + \mathbf{A}_\perp
    = -\nabla\phi + \nabla\times \mathbf{P}
\end{align}
where $\mathbf{A}_\parallel=-\nabla\phi$ and $\mathbf{A}_\perp=\nabla\times \mathbf{P}$.
Here we use $\mathbf{A}_\parallel$ and $\mathbf{A}_\perp$ to denote longitudinal fields and transverse fields, respectively.\footnote{It should be noted that the term longitudinal and transverse are first used in the context of plane wave, where $\nabla\times\mathbf{A}=0\rightarrow\mathbf{k}\times\mathbf{A}=0$ and $\nabla\cdot\mathbf{A}=0\rightarrow\mathbf{k}\cdot\mathbf{A}=0$, suggesting that the longitudinal (transverse) field is parallel (perpendicular) to the propagation direction in the Fourier space. Here we port the use of these terms to curl-free and divergence free in general cases, with a caveat that they do not have the same physical meaning as in plane wave.}  
Apparently $\nabla\times\mathbf{A}_\parallel=0$ and $\nabla\cdot\mathbf{A}_\perp=0$. 
Since a vector can be uniquely defined by its divergence $\nabla\cdot\mathbf{A}=s$ and curl $\nabla\times\mathbf{A}=\mathbf{c}$, taking the divergence and curl of the above, we have
\begin{align}
    \nabla\cdot\mathbf{A} &= -\nabla^2\phi = s \label{helmholtz_div} \\
    \nabla\times\mathbf{A} &= \nabla\times\nabla\times\mathbf{P} = \mathbf{c} \label{helmholtz_curl}
\end{align}
where $s$ acts as the source for Poisson equation, and $\mathbf{c}$ the source for $\mathbf{P}$.
We proceed the proof of Helmholtz theorem by showing the expression for $\phi$ and $\mathbf{A}$ that recover $s$ and $\mathbf{c}$.
Since the solution of $\mathbf{P}$ in (\ref{helmholtz_curl}) is not unique, we set $\nabla\cdot\mathbf{P}=0$, and (\ref{helmholtz_curl}) becomes vector Poisson equation with a unique solution. The solution to (\ref{helmholtz_div}) and (\ref{helmholtz_curl}) are then well known as \cite{jackson1999classical}
\begin{align}
    \phi(\mathbf{r}) &= \frac{1}{4\pi} \int \frac{s(\mathbf{r}')}{|\mathbf{r}-\mathbf{r}'|} d\mathbf{r}' \label{helmholtz_phi_recover_s} \\
    \mathbf{P}(\mathbf{r}) &= \frac{1}{4\pi} \int \frac{\mathbf{c}(\mathbf{r}')}{|\mathbf{r}-\mathbf{r}'|} d\mathbf{r}' \label{helmholtz_P_recover_c} .
\end{align}
The above can be proved by direct substitution of (\ref{helmholtz_phi_recover_s}) and (\ref{helmholtz_P_recover_c}) into (\ref{helmholtz_div}) and (\ref{helmholtz_curl}).

\subsection{Solution Space of Wave Equations}
Helmholtz theorem can be used to explain the null-space solution of electric field wave equation.
We consider the electric field wave equations in a homogeneous medium without sources 
\begin{align}
    \nabla\times\nabla\times\mathbf{E} = \omega^2\epsilon\mu \mathbf{E}  . \label{wave_eq_E_vacuum}
\end{align}
It can be formulated as an eigenvalue problem. 
According to the Helmholtz theorem, the solution to (\ref{wave_eq_E_vacuum}) must be either div-free, or curl-free.
Apparently curl-free solutions are the null-space solutions to (\ref{wave_eq_E_vacuum}), where the eigenvalues are zero.
On the other hand, div-free solutions are the non-null-space solutions. This is because when $\nabla\cdot\mathbf{E}=0$, we can add $-\nabla\nabla\cdot\mathbf{E}$ to the left-hand side of (\ref{wave_eq_E_vacuum}) without changing the equation. Then the left-hand side becomes $\nabla\times\nabla\times \mathbf{E} - \nabla\nabla\cdot\mathbf{E}=-\nabla^2\mathbf{E}$ and the Laplacian $\nabla^2$ is a negative definite operator. Thus the div-free solutions must have nonzero eigenvalues. 

\begin{figure}
    \centering
    \includegraphics[width=0.5\textwidth]{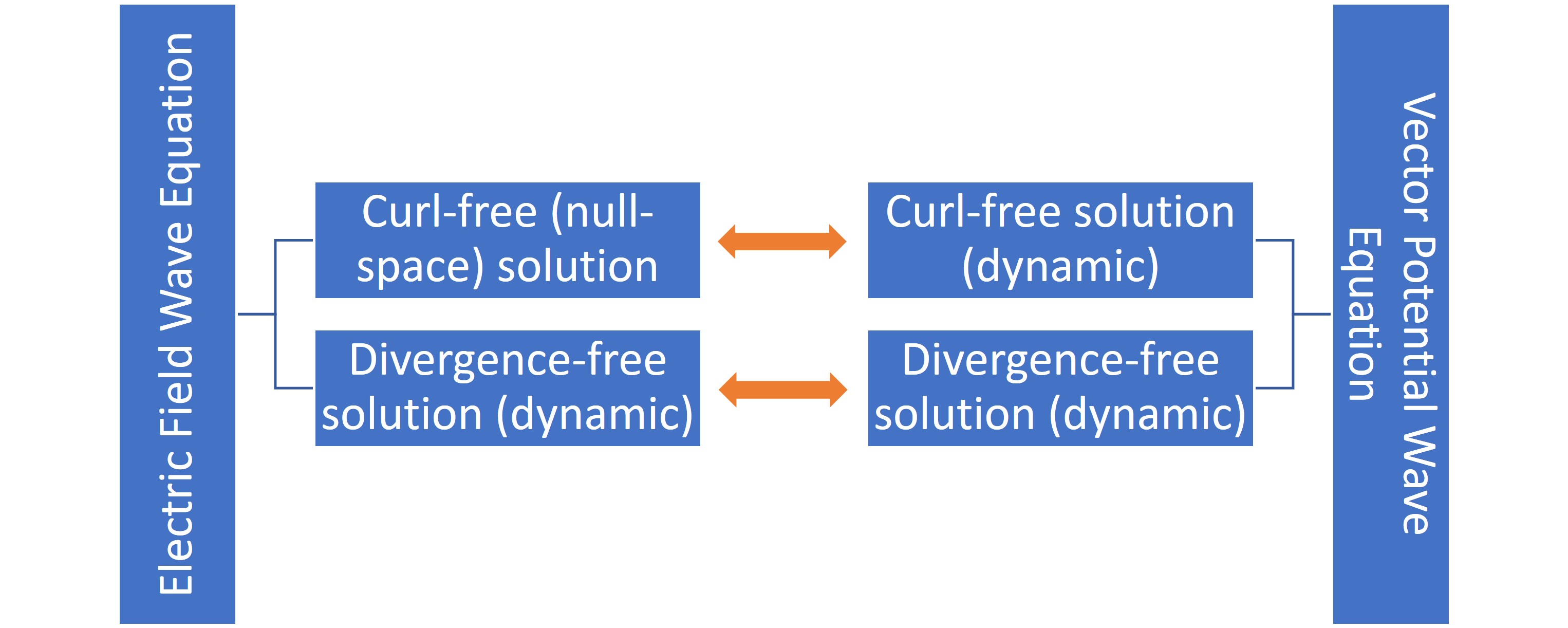}   
    \caption{Mapping between solution space of the electric field wave equation and vector potential wave equation.
    The double-headed arrow represents the same space spanned by the eigenvectors of both equations.}
    \label{fig:mapping}
\end{figure}

Helmholtz theorem also connects the solution to the vector potential wave equation with the solution to the electric field wave equation. 
We consider the vector potential wave equations with Lorenz gauge $\nabla\cdot\mathbf{A}=-\epsilon\mu\partial_t\Phi$ in the same homogeneous medium without sources 
\begin{align}
    \nabla\times\nabla\times \mathbf{A} - \nabla\nabla\cdot\mathbf{A} = \omega^2 \epsilon\mu \mathbf{A}  . \label{wave_eq_A_vacuum}
\end{align}
Both (\ref{wave_eq_E_vacuum}) and (\ref{wave_eq_A_vacuum}) have two families of solutions following the Helmholtz theorem.
The div-free solution to (\ref{wave_eq_E_vacuum}) must also be the div-free solution to (\ref{wave_eq_A_vacuum}), and vice versa. This is because the second term of (\ref{wave_eq_A_vacuum}) can be dropped in the case of div-free solution.
However, since null-space solution does not exist in (\ref{wave_eq_A_vacuum}) due to the negative-definiteness of Laplacian, the space spanned by the null-space solutions to (\ref{wave_eq_E_vacuum}) must be filled by the same number of non-null-space solutions to (\ref{wave_eq_A_vacuum}). 
Again due to the Helmholtz theorem, these solutions are curl-free solutions in (\ref{wave_eq_A_vacuum}) as well.
The mapping between the solution space of (\ref{wave_eq_E_vacuum}) and (\ref{wave_eq_A_vacuum}) is illustrated in Fig. \ref{fig:mapping}.
Apparently curl-free vector potential in electrodynamics leads to zero $\mathbf{E}$ and $\mathbf{H}$ even though $\mathbf{A}$ and $\Phi$ are nonzero. 
This also justifies the use of $\Phi=0$ gauge due to Lorenz gauge to remove the redundancy of the curl-free solutions in homogeneous case for many practical problems.

\subsection{A New Look of Helmholtz Theorem}
We can also prove Helmholtz theorem from a different perspective by looking at (\ref{wave_eq_E_vacuum}). 
We first prove that the curl operator is a self-adjoint operator \cite{chew1995waves} using inner product.
We begin with 
\begin{align}
    \langle \mathbf{A}^*, \overline{\bm{\mathcal{C}}} \mathbf{B} \rangle 
    &= \int_V d\mathbf{r} \,  \mathbf{A}^* \cdot
    \left(\nabla\times\mathbf{B} \right) & \nonumber \\
    &= \int_V d\mathbf{r} \,  \mathbf{B} \cdot
    \left(\nabla\times\mathbf{A}^* \right)
    - \int_S d\mathbf{r}  \, \hat{n} \cdot (\mathbf{A}^* \times \mathbf{B})
\end{align}
where $\overline{\bm{\mathcal{C}}}$ denotes the curl operator.
With the boundary condition that $\hat{n}\times\mathbf{A}=0$ and $\hat{n}\times\mathbf{B}=0$,\footnote{The rationale of this boundary condition is explained in the next section.}
and the surface integral term vanishes, and hence the curl operator is self-adjoint for fields of the prescribed boundary conditions. With the same approach, it can also be proved that $\nabla\times\nabla\times$ is a self-adjoint operator \cite{roth2022theory}.
Thus the solution vectors to (\ref{wave_eq_E_vacuum}) form a complete basis. 

The proceeding proof also suggests that the curl operator has the same null-space as its power. In general when an operator maps a domain space function (or a row space vector, if the operator is a finite-sized matrix) into its range space (or column space), the new function may contain null-space component. 
However, the self-adjointness and symmetry of the curl operators ensure that its domain space and range space are the same, and its (right-)null-space and left-null-space are the same. Thus the curl operator maps a domain space function into the same space. Hence it will not generate a null-space component. 

Now we are ready to prove the Helmholtz theorem by analyzing (\ref{wave_eq_E_vacuum}). 
We only consider the non-trivial solution to (\ref{wave_eq_E_vacuum}) where $\mathbf{E}\neq 0$ almost everywhere. 
Taking the divergence of (\ref{wave_eq_E_vacuum}), we have $\omega^2\nabla\cdot\mathbf{E}=0$. 
It has three possible classes of solutions: (i) $\omega=0$ and $\nabla\cdot\mathbf{E}\neq 0$; (ii) $\omega\neq0$ and $\nabla\cdot\mathbf{E}=0$; and (iii) $\omega=0$ and $\nabla\cdot\mathbf{E}=0$.
We can rule out (iii) because it would make (\ref{wave_eq_E_vacuum}) the same as (\ref{wave_eq_A_vacuum}), and $\omega=0$ solution cannot exist in (\ref{wave_eq_A_vacuum}). This is because $\nabla\times\nabla\times \mathbf{A} - \nabla\nabla\cdot\mathbf{A}=-\nabla^2\mathbf{A}$ and the Laplacian $\nabla^2$ is a negative definite operator.
For (i), apparently $\nabla\times\nabla\times\mathbf{E}=0$, and in turn $\nabla\times\mathbf{E}=0$.
For (ii), we have $\nabla\times\mathbf{E}\neq 0$. Concluding the above, we can clearly see two classes of solution to (\ref{wave_eq_E_vacuum}) similar to the Helmholtz theorem, where
\begin{enumerate}
    \item $\nabla\times\mathbf{E}=0$, $\nabla\cdot\mathbf{E}\neq 0$;
    \item $\nabla\cdot\mathbf{E}= 0$, $\nabla\times\mathbf{E}\neq0$.
\end{enumerate}
Due to the self-adjointness of $\nabla\times\nabla\times$ operator, the solution to (\ref{wave_eq_E_vacuum}) form a complete basis to the entire space. Hence any function satisfying the proposed boundary condition can be separated into two classes.
This completes the proof of Helmholtz theorem. 
Note that analyzing (\ref{wave_eq_A_vacuum}) also leads to the same two families of solutions described above.
This will be shown in greater details similar to the inhomogeneous media case next.

\subsection{Motivation of Generalization}
As shown above, the Helmholtz theorem works well in explaining the solution space of the electric field wave equation and vector potential wave equation with Lorenz gauge in homogeneous medium.
However, in inhomogeneous dispersionless media, it is suggested that generalized Lorenz gauge should be used instead \cite{chew2014vector}. 
Thus, to better explain the eigenmodes found by potential-based formulation, it would be ideal if Helmholtz theorem can be generalized in a way that $\nabla\cdot\epsilon(\mathbf{r})\mathbf{A}_{\epsilon\perp}(\mathbf{r})=0$,\footnote{Position-dependent $\epsilon(\mathbf{r})$ is assumed from now on. For simplicity, we only write $\epsilon$. }
where $\mathbf{A}_{\epsilon\perp}$ denotes the div-$\epsilon$-free (or generalized transverse) fields.
Considering this, we seek to prove the following generalized Helmholtz decomposition:
\begin{displayquote}
\textit{The vector field $\mathbf{A}$ given by the Helmholtz wave equation in source-free inhomogeneous media can be decomposed into two components $\mathbf{A}_{\epsilon\perp}$ and $\mathbf{A}_\parallel$, where the div-$\epsilon$-free (or generalized transverse) fields satisfy $\nabla\cdot\epsilon\mathbf{A}_{\epsilon\perp}=0$ and the curl-free (or longitudinal) fields satisfy $\nabla\times\mathbf{A}_\parallel=0$.} 
\end{displayquote}

In the remainder of this work, we use solution of the wave equation for vector potential $\mathbf{A}$ to arrive at the generalized Helmholtz decomposition. 
A modified generalized Lorenz gauge $\nabla\cdot\epsilon\mathbf{A}=-\epsilon_0^2\mu_0\partial_t\Phi$ is used.
We follow matrix theory generally in the proof. It will be shown that the vector potential wave equation has two classes of solutions. They degenerate into the Helmholtz decomposition in free space.
Thus the vector potential wave equation for inhomogeneous media can be used to motivate the generalized Helmholtz decomposition.
We also note that there is no null space in the resulting generalized eigenvalue problem.

\section{Proof of Orthogonality}\label{section:orthogonality}
We start from proving the orthogonality between the two families of fields proposed in the generalized Helmholtz decomposition, assuming the decomposition does exist. 
We leave it to the next section to demonstrate the existence of the decomposition through numerical examples. 

\begin{figure}
    \centering
    \includegraphics[width=0.4\textwidth]{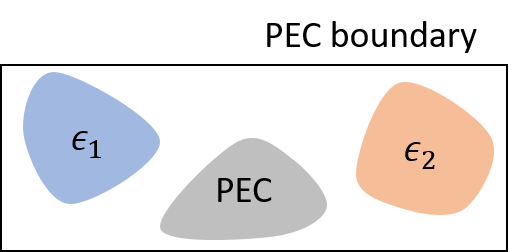}   
    \caption{Schematic of the region of interest.}
    \label{fig:schematic}
\end{figure}

We consider the space terminated by continuous perfect electric conductor (PEC) boundary, filled with inhomogeneous media (Fig. \ref{fig:schematic}). 
Because of the gauge $\nabla\cdot\epsilon\mathbf{A}=-\epsilon_0^2\mu_0\partial_t\Phi$, the boundary condition for $\nabla\cdot\epsilon\mathbf{A}$ is actually that for $\Phi$ on a PEC, which is $\Phi$ equals a constant voltage \cite{vico2016decoupled,roth2022theory}.
If there is only one PEC, we can choose this as our reference.
Thus
\begin{align}
    \hat{n} \times \mathbf{A} = 0
\end{align}
which is consistent with $\hat{n} \times \mathbf{E} = 0$, and
\begin{align}
    \nabla\cdot\epsilon\mathbf{A}=0 .
\end{align}
The latter is equivalent to $\Phi=0$ on the boundary when the PEC has a constant voltage of zero, i.e. ``grounded".

If there are additional PECs inside the domain, they may not be grounded. In this case, $\nabla\cdot\epsilon\mathbf{A}=\Phi$ is equal to a constant voltage on each conductor. Additionally, charge neutrality constraint can be enforced on each conductor, which requires \cite{roth2022theory,vico2016decoupled}
\begin{align}
    \int_S d\mathbf{r} \, \hat{n}\cdot\epsilon\mathbf{A} \; = 0. \label{charge_neutral}
\end{align} 
where the surface integral is over the surface of each PEC.

Next, we proceed to show the proof of orthogonality first in continuous space, followed by a similar procedure in discrete space. 
Different definitions of orthogonality will be discussed.
It will also be shown that discrete approximation of Maxwell’s equations is homomorphic to the continuum case if done properly.

\subsection{Continuous Space}\label{subsection:orthogonality_continuous}
Wave equation is usually the starting point of many $\mathbf{A}$-$\Phi$ based solver. We start the mathematical proof of mode orthogonality by defining proper operator and vector space. 
We consider the wave equation for vector potential $\mathbf{A}$ in lossy dispersionless inhomogeneous isotropic media\footnote{To satisfy the Kramers–Kronig relations, this assumption is only valid over a narrow bandwidth.} with modified generalized Lorenz gauge \cite{chew2014vector}
\begin{align}
    \nabla\times\frac{1}{\mu}\nabla\times \mathbf{A} - \frac{1}{\epsilon_0^2\mu_0}\epsilon\nabla\nabla\cdot\epsilon \mathbf{A} = \omega^2 \epsilon \mathbf{A} . \label{wave_eq_A_evp}
\end{align}

Let
\begin{align}
    \overline{\bm{\mathcal{W}}}_1 &= \nabla\times\frac{1}{\mu}\nabla\times \label{operator_W1} \\
    \overline{\bm{\mathcal{W}}}_2 &= - \frac{1}{\epsilon_0^2\mu_0}\epsilon\nabla\nabla\cdot\epsilon \label{operator_W2} \\
    \overline{\bm{\mathcal{L}}} &= \overline{\bm{\mathcal{W}}}_1 + \overline{\bm{\mathcal{W}}}_2 \label{operator_L} \\
    \overline{\bm{\mathcal{U}}} &= \epsilon \label{operator_U} \\
    \lambda &= \omega^2 .
\end{align}
Then (\ref{wave_eq_A_evp}) can be written abstractly and concisely as
\begin{align}
    (\overline{\bm{\mathcal{W}}}_1 + \overline{\bm{\mathcal{W}}}_2) \mathbf{A} = \lambda \bm{\overline{\mathcal{U}}} \mathbf{A} \label{wave_eq_A_evp_abstract}
\end{align}
where the operator notation is defined as a differential operator acting on a function $\mathbf{A}$. 
Note that we use boldface calligraphic font with an overbar to denote a Hilbert space operator. The function $\mathbf{A}$ in the abovementioned space is in turn square integrable \cite{axelsson2001finite}. 


It has been proved that the $\overline{\bm{\mathcal{L}}}$ operator is self-adjoint in lossless media given the boundary conditions defined in the beginning of this section \cite{roth2022theory}. Here we investigate the general case with lossy media.
Since the $\overline{\bm{\mathcal{L}}}$ operator consists of two terms $\overline{\bm{\mathcal{W}}}_1$ and $\overline{\bm{\mathcal{W}}}_2$, it is clearer to look at the two terms separately \cite{roth2022theory}.

We first define $\mathbf{A}_{\epsilon\perp}$ to be the solution to the eigenvalue problem $\overline{\bm{\mathcal{W}}}_1 \mathbf{A}_{\epsilon\perp} = \lambda_{\epsilon\perp} \overline{\bm{\mathcal{U}}} \mathbf{A}_{\epsilon\perp}$ with eigenvalue $\lambda_{\epsilon\perp}$. Here we only study the solutions with nonzero eigenvalues. This gives us, from the definition above,
\begin{align}
    \nabla\times\frac{1}{\mu}\nabla\times \mathbf{A}_{\epsilon\perp} = \lambda_{\epsilon\perp} \epsilon \mathbf{A}_{\epsilon\perp} . \label{evp_curl_curl}
\end{align}
Taking divergence of the above equation gives
\begin{align}
    0 = \lambda_{\epsilon\perp} \nabla\cdot\epsilon\mathbf{A}_{\epsilon\perp} . \label{null_A_perp}
\end{align}
Thus $\mathbf{A}_{\epsilon\perp}$ is in the null space of the $\overline{\bm{\mathcal{W}}}_2$ operator defined in (\ref{operator_W2}), and it is also a solution to the original eigenvalue problem (\ref{wave_eq_A_evp_abstract}) with eigenvalue $\lambda_{\epsilon\perp}$.

Similarly, we define $\mathbf{A}_\parallel$ to be the solution to $\overline{\bm{\mathcal{W}}}_2 \mathbf{A}_\parallel = \lambda_\parallel \overline{\bm{\mathcal{U}}} \mathbf{A}_\parallel$ with nonzero eigenvalue $\lambda_\parallel$, or more explicitly,
\begin{align}
    - \frac{1}{\epsilon_0^2\mu_0} \epsilon\nabla\nabla\cdot\epsilon \mathbf{A}_\parallel = \lambda_\parallel \epsilon \mathbf{A}_\parallel . \label{evp_grad_div}
\end{align}
Dividing the above by $\epsilon$ and taking curl of the resulting equation yields
\begin{align}
    0 = \lambda_\parallel \nabla\times \mathbf{A}_\parallel . \label{null_A_parallel}
\end{align}
Thus $\mathbf{A}_\parallel$ is in the null space of the $\overline{\bm{\mathcal{W}}}_1$ operator defined in (\ref{operator_W1}), and it is also a solution to the original eigenvalue problem (\ref{wave_eq_A_evp_abstract}) with eigenvalue $\lambda_\parallel$.

Next, we will manipulate the above equations, and use (\ref{null_A_perp}) and (\ref{null_A_parallel}) to prove orthogonality of two sets of modes.

\subsubsection{Proof 1}
To begin, we take the left dot product of (\ref{evp_grad_div}) by $\mathbf{A}_{\epsilon\perp}$, and integrate it over the space filled with dielectrics to yield
\begin{align}
    - \frac{1}{\epsilon_0^2\mu_0} \int_V d\mathbf{r} \left(  \mathbf{A}_{\epsilon\perp} \cdot \epsilon\nabla\nabla\cdot\epsilon \mathbf{A}_\parallel \right) = \int_V d\mathbf{r} \left( \mathbf{A}_{\epsilon\perp} \cdot \lambda_\parallel \epsilon \mathbf{A}_\parallel \right) . \label{integration_p1}
\end{align}
Using integration by parts, the left-hand side becomes
\begin{align}
    \frac{1}{\epsilon_0^2\mu_0} \int_V d\mathbf{r} \,  (\nabla\cdot\epsilon\mathbf{A}_{\epsilon\perp}) (\nabla\cdot\epsilon\mathbf{A}_\parallel) & \nonumber \\
    - \frac{1}{\epsilon_0^2\mu_0} \int_S d\mathbf{r} \, \hat{n} \cdot & \left[  (\nabla\cdot\epsilon\mathbf{A}_\parallel)\epsilon\mathbf{A}_{\epsilon\perp} \right]  \label{integration_by_part_p1}
\end{align}
where the surface integral is over the PEC surfaces. 
From (\ref{null_A_perp}), the first term vanishes. At the grounded PEC boundary, $\nabla\cdot\epsilon\mathbf{A}_\parallel=0$, and then the second term also vanishes. 
At the PEC not connected to the ground, the second term becomes
\begin{align}
    - \frac{1}{\epsilon_0^2\mu_0} (\nabla\cdot\epsilon\mathbf{A}_\parallel) \int_S d\mathbf{r} \, \hat{n} \cdot  \epsilon\mathbf{A}_{\epsilon\perp} = 0 
\end{align}
where the divergence term is taken outside the integral because $\Phi$ is equal to a constant voltage on the conductor, and the integral vanishes because of the charge neutrality constraint (\ref{charge_neutral}).
Thus
\begin{align}
    \langle \mathbf{A}_{\epsilon\perp}, \epsilon \mathbf{A}_\parallel \rangle = 0 \label{orthogonality_p1}
\end{align}
as long as $\lambda_\parallel \neq 0$, where $\langle\mathbf{F},\mathbf{G}\rangle=\int\mathbf{F}\cdot\mathbf{G} \;d\mathbf{r}$ is defined as the reaction inner product by Rumsey \cite{rumsey1954reaction}. 
In other words, the two sets of modes $\mathbf{A}_{\epsilon\perp}$ and $\mathbf{A}_\parallel$ are $\epsilon$-orthogonal to each other.

\subsubsection{Proof 2}
Now, taking the left dot product of (\ref{evp_curl_curl}) with $\mathbf{A}_\parallel$, and integrating it over space yields
\begin{align}
    \int_V d\mathbf{r} \left(  \mathbf{A}_\parallel \cdot \nabla\times\frac{1}{\mu}\nabla\times \mathbf{A}_{\epsilon\perp} \right) 
    = \int_V d\mathbf{r} \left( \mathbf{A}_\parallel \cdot \lambda_{\epsilon\perp} \epsilon \mathbf{A}_{\epsilon\perp} \right) . \label{integration_p2}
\end{align}
The right-hand side is denoted $\lambda_{\epsilon\perp} \langle \mathbf{A}_\parallel, \epsilon \mathbf{A}_{\epsilon\perp} \rangle$. Using integration by parts, the left-hand side becomes
\begin{align}
    \int_V d\mathbf{r} \,  \left(\frac{1}{\mu}\nabla\times\mathbf{A}_{\epsilon\perp}\right) \cdot
    \left(\nabla\times\mathbf{A}_\parallel \right) & \nonumber \\
    + \int_S d\mathbf{r}  \, \hat{n} \cdot & \left[  \left(\frac{1}{\mu}\nabla\times\mathbf{A}_{\epsilon\perp}\right)  \times\mathbf{A}_\parallel  \right] . \label{integration_by_part_p2}
\end{align}
From (\ref{null_A_parallel}), the first term vanishes. Implementing the boundary condition that $\hat{n}\times\mathbf{A}_\parallel=0$, the second term also vanishes. Thus
\begin{align}
    \langle \mathbf{A}_\parallel, \epsilon \mathbf{A}_{\epsilon\perp} \rangle = 0 \label{orthogonality_p2}
\end{align}
as long as $\lambda_{\epsilon\perp} \neq 0$. In other words, the two sets of modes $\mathbf{A}_\parallel$ and $\mathbf{A}_{\epsilon\perp}$ are $\epsilon$-orthogonal to each other.

In the above derivation, if we replace $\mathbf{A}_\parallel$ by $\mathbf{A}_\parallel^*$, the derivation is still valid. That is 
\begin{align}
    \langle \mathbf{A}_\parallel^*, \epsilon \mathbf{A}_{\epsilon\perp} \rangle = 0 . \label{orthogonality_p2_cc}
\end{align}
However, the same token does not apply to \textit{Proof 1} in \ref{subsection:orthogonality_continuous}.
This may motivate new definition of inner product in the problems concerning potential-based formulation.

\subsection{Discrete Space}\label{subsection:orthogonality_discrete}

The problem in the previous subsection is homomorphic to numerical linear algebra if we find the matrix representation of the operators using subspace projection methods \cite{chew604}, including finite difference method \cite{taflove2005computational}, finite element method \cite{jin2015finite}, or discrete exterior calculus \cite{chen2017electromagnetic}. 
These methods should be chosen properly to preserve certain properties of the continuum calculus, i.e. $\nabla\cdot(\nabla\times\mathbf{A})=0$ and $\nabla\times\nabla f=0$. 
In this work we use finite difference method. 
After discretization, we have
\begin{align}
    \nabla\times \mathbf{A} &\Rightarrow \overline{\mathbf{C}}_1 \cdot \mathsf{A}_1 \label{discrete_1} \\
    \nabla\times\nabla\times \mathbf{A} &\Rightarrow \overline{\mathbf{C}}_2 \cdot \overline{\mathbf{C}}_1 \cdot \mathsf{A}_1 \label{discrete_2} \\
    \nabla\cdot\epsilon\mathbf{A} &\Rightarrow \overline{\mathbf{D}}_1 \cdot \mathsf{A}_1 \label{discrete_3} \\
    \epsilon\nabla f &\Rightarrow \overline{\mathbf{E}}_1 \cdot \mathsf{f}_1 \label{discrete_4} \\
    \epsilon &\Rightarrow \overline{\mathbf{U}} \label{discrete_5} 
\end{align}
where the operators on the LHS are in the continuous Hilbert space, while the RHS are the linear algebra approximation of the continuum space.
Here we use boldface font with overbar to denote finite size matrices, and sans serif typestyle to denote finite length vectors stored as 1D arrays in computers. 

Due to the discretization, each operator may have different representations depending on the space it operates on. To use finite difference method, Yee grid is constructed with both primal and dual grids \cite{yee1966numerical,taflove2005computational}. On each grid, we place a 3D vector on the face of the cube, and a scalar on the center of the cube. Details of the discretization scheme can be found in the Appendix A. 

In (\ref{discrete_1}) to (\ref{discrete_5}), the discrete operators with subscript $1$ denotes operations on the primal mesh, while subscript $2$ denotes operations on the dual mesh. 
The discrete curl operator $\overline{\mathbf{C}}_1$ acts on a vector on the primal mesh, resulting in a vector on the dual mesh, while $\overline{\mathbf{C}}_2$ acts on a vector on the dual mesh and results in a vector on the primal mesh. 
It can be shown that these two discrete curl operators are transpose to each other $\overline{\mathbf{C}}_1=\overline{\mathbf{C}}_2^T$, where $^T$ denotes matrix transpose. 
Also, it is easy to prove that the discrete divergence operator and the negative of discrete gradient operator are transpose to each other $\overline{\mathbf{D}}_{1,2}^T=-\overline{\mathbf{E}}_{1,2}$.
The $\overline{\mathbf{U}}$ operator is a square matrix whose dimension depends on the vector it operates on.
The following vector identities are preserved after discretization using finite difference method \cite{chew1994electromagnetic,teixeira1999lattice,weiland1996time}
\begin{align}
    \nabla\cdot(\nabla\times\mathbf{A})=0 &\Rightarrow \overline{\mathbf{D}}_2\cdot\overline{\mathbf{U}}^{-1}_2\cdot\overline{\mathbf{C}}_1\cdot\mathsf{A}_1=0 \nonumber \\
    & \mathrm{and} \;
    \overline{\mathbf{D}}_1\cdot\overline{\mathbf{U}}^{-1}_1\cdot\overline{\mathbf{C}}_2\cdot\mathsf{A}_2=0 \label{vec_id_1_matrix} \\
    \nabla\times\nabla f=0 &\Rightarrow \overline{\mathbf{C}}_1\cdot\overline{\mathbf{U}}^{-1}_1\cdot\overline{\mathbf{E}}_1\cdot \mathsf{f}_1 \nonumber \\ 
    &= -\overline{\mathbf{C}}_1\cdot\overline{\mathbf{U}}^{-1}_1\cdot\overline{\mathbf{D}}_1^T\cdot \mathsf{f}_1 = 0 . \label{vec_id_2_matrix}
\end{align}
Without loss of generality, we let $\mu=1$.\footnote{It can be seen that the proof is still valid when $\mu$ is inhomogeneous, anisotropic, or complex.} 
Also we denote $\alpha=1/\epsilon_0^2\mu_0$. We assume the solution vector is on the primal mesh, and thus the subscript $1$ is omitted. 
Then (\ref{wave_eq_A_evp_abstract}) becomes
\begin{align}
    \overline{\mathbf{C}}_2\cdot\overline{\mathbf{C}}_1\cdot\mathsf{A} - \alpha\overline{\mathbf{E}}_1\cdot\overline{\mathbf{D}}_1\cdot\mathsf{A}  = \lambda\overline{\mathbf{U}}_1\cdot \mathsf{A} .
\end{align}

We break the above into two eigenvalue problems corresponding to the two operators on the left-hand side
\begin{align}
    \overline{\mathbf{C}}_2\cdot\overline{\mathbf{C}}_1\cdot\mathsf{A}_{\epsilon\perp} &= \lambda_{\epsilon\perp} \overline{\mathbf{U}}_1\cdot \mathsf{A}_{\epsilon\perp} \label{evp_curl_curl_matrix} \\
    -\alpha\overline{\mathbf{E}}_1\cdot\overline{\mathbf{D}}_1\cdot\mathsf{A}_\parallel  &= \lambda_\parallel \overline{\mathbf{U}}_1\cdot \mathsf{A}_\parallel . \label{evp_grad_div_matrix}
\end{align}

Taking the left dot product of (\ref{evp_curl_curl_matrix}) with $\overline{\mathbf{D}}_1\cdot\overline{\mathbf{U}}_1^{-1}$, we get
\begin{align}
    \overline{\mathbf{D}}_1\cdot\overline{\mathbf{U}}_1^{-1}\cdot\overline{\mathbf{C}}_2\cdot\overline{\mathbf{C}}_1\cdot\mathsf{A}_{\epsilon\perp} &= \lambda_{\epsilon\perp} \overline{\mathbf{D}}_1\cdot\overline{\mathbf{U}}_1^{-1}\cdot \overline{\mathbf{U}}_1\cdot \mathsf{A}_{\epsilon\perp} .
\end{align}
Due to (\ref{vec_id_1_matrix}), LHS of the above vanishes. Thus
\begin{align}
    \overline{\mathbf{D}}_1\cdot\mathsf{A}_{\epsilon\perp}=0 . \label{null_A_perp_matrix}
\end{align}

Taking the left dot product of (\ref{evp_grad_div_matrix}) with $\overline{\mathbf{C}}_1\cdot\overline{\mathbf{U}}_1^{-1}$, we get
\begin{align}
    -\alpha\overline{\mathbf{C}}_1\cdot\overline{\mathbf{U}}_1^{-1}\cdot\overline{\mathbf{E}}_1\cdot\overline{\mathbf{D}}_1\cdot\mathsf{A}_\parallel  &= \lambda_\parallel \overline{\mathbf{C}}_1\cdot\overline{\mathbf{U}}_1^{-1}\cdot\overline{\mathbf{U}}_1\cdot \mathsf{A}_\parallel .
\end{align}
Due to (\ref{vec_id_2_matrix}), the LHS of the above vanishes. Thus
\begin{align}
    \overline{\mathbf{C}}_1\cdot\mathsf{A}_\parallel=0 . \label{null_A_parallel_matrix}
\end{align}

\subsubsection{Proof 1}
Taking the left dot product of (\ref{evp_grad_div_matrix}) with $\mathsf{A}_{\epsilon\perp}^T$ and using the symmetry property between divergence and gradient operators, we have
\begin{align}
    \alpha\mathsf{A}_{\epsilon\perp}^T\cdot \overline{\mathbf{D}}_1^T\cdot\overline{\mathbf{D}}_1\cdot\mathsf{A}_\parallel  &= \lambda_\parallel \mathsf{A}_{\epsilon\perp}^T\cdot \overline{\mathbf{U}}_1\cdot \mathsf{A}_\parallel .
\end{align}
Due to (\ref{null_A_perp_matrix}) and its transpose, LHS of the above vanishes. Thus
\begin{align}
    \mathsf{A}_{\epsilon\perp}^T\cdot \overline{\mathbf{U}}\cdot \mathsf{A}_\parallel = 0 .
\end{align}
which is analogous to (\ref{orthogonality_p1}).

\subsubsection{Proof 2}
Now, taking the left dot product of  (\ref{evp_curl_curl_matrix}) with $\mathsf{A}_\parallel^T$, where $^\dagger$ denotes conjugate transpose, and using the symmetry property of curl operator, we have
\begin{align}
    \mathsf{A}_\parallel^T\cdot \overline{\mathbf{C}}_1^T\cdot\overline{\mathbf{C}}_1\cdot\mathsf{A}_{\epsilon\perp} &= \lambda_{\epsilon\perp} \mathsf{A}_\parallel^T\cdot  \overline{\mathbf{U}}_1\cdot \mathsf{A}_{\epsilon\perp} .
\end{align}
Due to (\ref{null_A_parallel_matrix}) and its transpose, LHS of the above vanishes. Thus
\begin{align}
    \mathsf{A}_\parallel^T\cdot \overline{\mathbf{U}}_1\cdot \mathsf{A}_{\epsilon\perp} = 0
\end{align}
which is homomorphic to (\ref{orthogonality_p2}).
However, we can also left dot product (\ref{evp_curl_curl_matrix}) by $\mathsf{A}_\parallel^\dagger$, and get
\begin{align}
    \mathsf{A}_\parallel^\dagger\cdot \overline{\mathbf{C}}_1^T\cdot\overline{\mathbf{C}}_1\cdot\mathsf{A}_{\epsilon\perp} &= \lambda_{\epsilon\perp} \mathsf{A}_\parallel^\dagger\cdot  \overline{\mathbf{U}}_1\cdot \mathsf{A}_{\epsilon\perp} .
\end{align}
Due to (\ref{null_A_parallel_matrix}) and its Hermitian conjugate, and the realness of the curl operator under finite difference method, LHS of the above vanishes. Thus
\begin{align}
    \mathsf{A}_\parallel^\dagger\cdot \overline{\mathbf{U}}\cdot \mathsf{A}_{\epsilon\perp} = 0
\end{align}
which is analogous to (\ref{orthogonality_p2_cc}).
The above shows that the finite difference approximation of Maxwell's equations is homomorphic to the continuum case if done properly.

\subsection{Anisotropic Case}\label{subsection:anisotropic}
The mathematical proof in the previous two subsections still applies in anisotropic media. Consider the wave equation for vector potential $\mathbf{A}$ in dispersionless inhomogeneous anisotropic media with modified generalized Lorenz gauge in frequency domain \cite{chew2014vector}
\begin{align}
    \nabla\times \overline{\bm{\mu}}^{-1}\nabla\times \mathbf{A} - \frac{1}{\epsilon_0^2\mu_0}\overline{\bm{\epsilon}}\cdot\nabla \left( \nabla\cdot\overline{\bm{\epsilon}} \cdot \mathbf{A} \right) = \omega^2 \overline{\bm{\epsilon}} \cdot \mathbf{A} . \label{wave_eq_A_evp_aniso}
\end{align}
The mathematical proof of mode orthogonality follows exactly as Section \ref{subsection:orthogonality_continuous} for continuous space and as Section \ref{subsection:orthogonality_discrete} for discrete space,
with the requirement that
\begin{align}
    \overline{\bm{\epsilon}} = \overline{\bm{\epsilon}}^T
\end{align}
which implies that the material is reciprocal.

\section{Numerical Results}\label{section:numerical}

\subsection{Numerical Demonstration of Completeness}\label{subsection:numerical_completeness}

In this section, we first demonstrate that the two sets of modes given in the previous section are complete in the discrete space.
We solve (\ref{wave_eq_A_evp}) numerically.
Finite difference method is used corresponding to the discrete space proof in Section \ref{section:orthogonality}. 
Uniform discretization is used, and the discretization scheme follows \cite{chew1994electromagnetic,ryu2016finite}, as described in Appendix \ref{appendix:discrete}.

\begin{figure}
    \centering
    \includegraphics[width=0.5\textwidth]{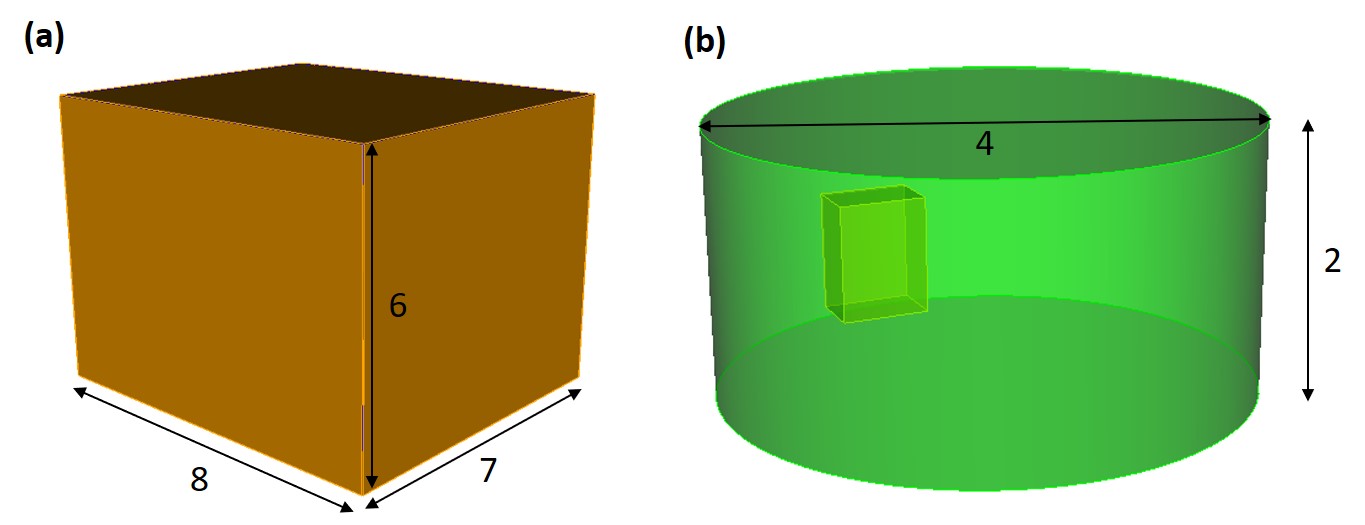}  
    \caption{Geometries of the numerical examples with dimensions.}
    \label{fig:geometry}
\end{figure}

\begin{figure}
    \centering
    \includegraphics[width=0.5\textwidth]{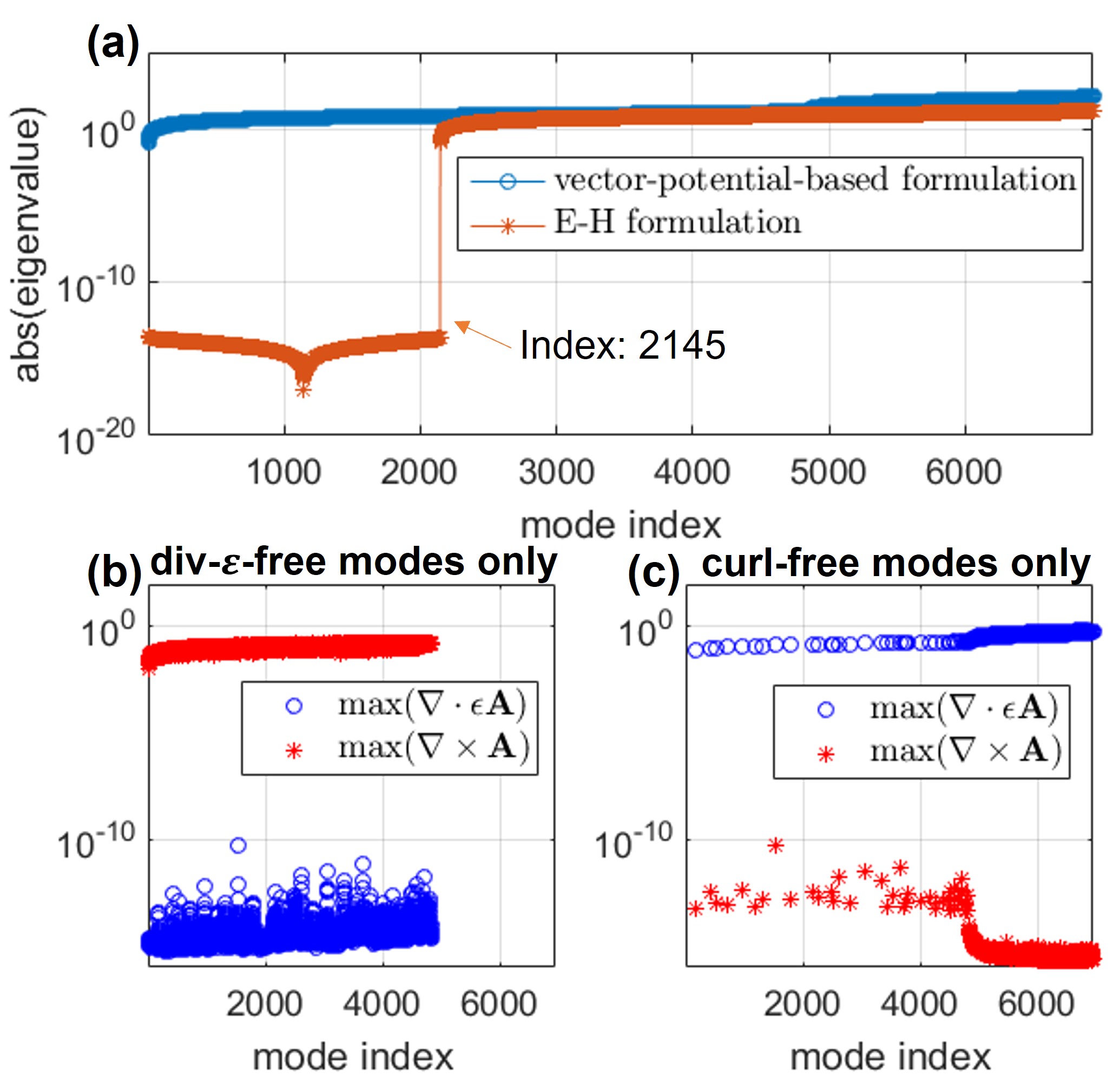}   
    \caption{Two set of modes found in the geometry described by Fig. \ref{fig:geometry}(a) can be clearly distinguished. 
    (a) The absolute value of eigenvalues are shown for both potential-based formulation and $\mathbf{E}$-$\mathbf{H}$ formulation. It can be seen clearly that the potential-based formulation does not have null space.
    For the potential-based formulation, 
    only the div-$\epsilon$-free (or generalized transverse) modes are shown in (b), and 
    only the curl-free (or longitudinal) modes are shown in (c).
    Totally there are 4793 div-$\epsilon$-free modes in (b) and 2145 curl-free modes in (c). 
    The latter is consistent with the number of null-space eigenvalues with $\mathbf{E}$-$\mathbf{H}$ formulation in (a).
    The total number of both families of modes is equal to the dimension of the eigenvalue problem.
    The indices of curl-free modes and div-$\epsilon$-free modes are skipped in (b) and (c), respectively. 
    It can be seen that the div-$\epsilon$-free modes satisfy $\nabla\cdot\epsilon\mathbf{A}=0$,
    while the curl-free modes satisfy $\nabla\times\mathbf{A}=0$.
    Blue open circles: for each mode, take $\nabla\cdot\epsilon\mathbf{A}$, and record the maximum value of abs($\nabla\cdot\epsilon\mathbf{A}$) in the entire grid.
    Red stars: for each mode, take $\nabla\times\mathbf{A}$, and record the maximum magnitude of all components in the entire grid.}
    \label{fig:numerical_proof_completeness_homo}
\end{figure}

For an empty or homogeneous structure, the proposed generalized Helmholtz decomposition degenerates into the usual Helmholtz decomposition. To demonstrate it, we consider a rectangular $7\times8\times6$ cavity homogeneously filled with a $\epsilon_r=3$ material, as shown in Fig. \ref{fig:geometry}(a). Discretization step is $0.5$, resulting in $6938 \times 6938$ matrices. We use natural unit so that $\epsilon_0=\mu_0=1$. 
After solving the generalized eigenvalue problem, we take divergence and curl on each mode to identify the two classes of modes. 
The results are displayed in Fig. \ref{fig:numerical_proof_completeness_homo}. 
Note that the indices of curl-free modes and div-$\epsilon$-free modes are skipped in (b) and (c), respectively. 
The total number of div-$\epsilon$-free and curl-free modes equals to the dimension of the eigenvalue problem.
Thus it clearly shows two distinct groups of modes corresponding to the Helmholtz decomposition, which complete the discrete space, and are consistent across field-based and potential-based formulations.

We then consider a more general example with anisotropic material. As shown in \ref{fig:geometry}(b), we construct a cylindrical cavity with radius $r=2$ and height $h=2$. It is filled with a block of anisotropic material centered at $(1,0.4,0.2)$ with dimension $0.6, 0.4, 0.8$ in $x,y,z$ directions, respectively. Discretization step is $0.2$, resulting in $5822 \times 5822$ matrices. The permittivity tensor of the anisotropic material is
\begin{align}
    \overline{\bm{\epsilon}} = 
    \begin{pmatrix}
        2 & 0 & 1+3i \\
        0 & 3+0.2i & 0 \\
        1+3i & 0 & 2
    \end{pmatrix}
\end{align}
which satisfies reciprocity. The matrix of anisotropic permittivity is treated the same way as \cite{rumpf2014finite}. The discretized operators satisfy the discrete vector identities (\ref{vec_id_1_matrix}) and (\ref{vec_id_2_matrix}). 

\begin{figure}
    \centering
    \includegraphics[width=0.5\textwidth]{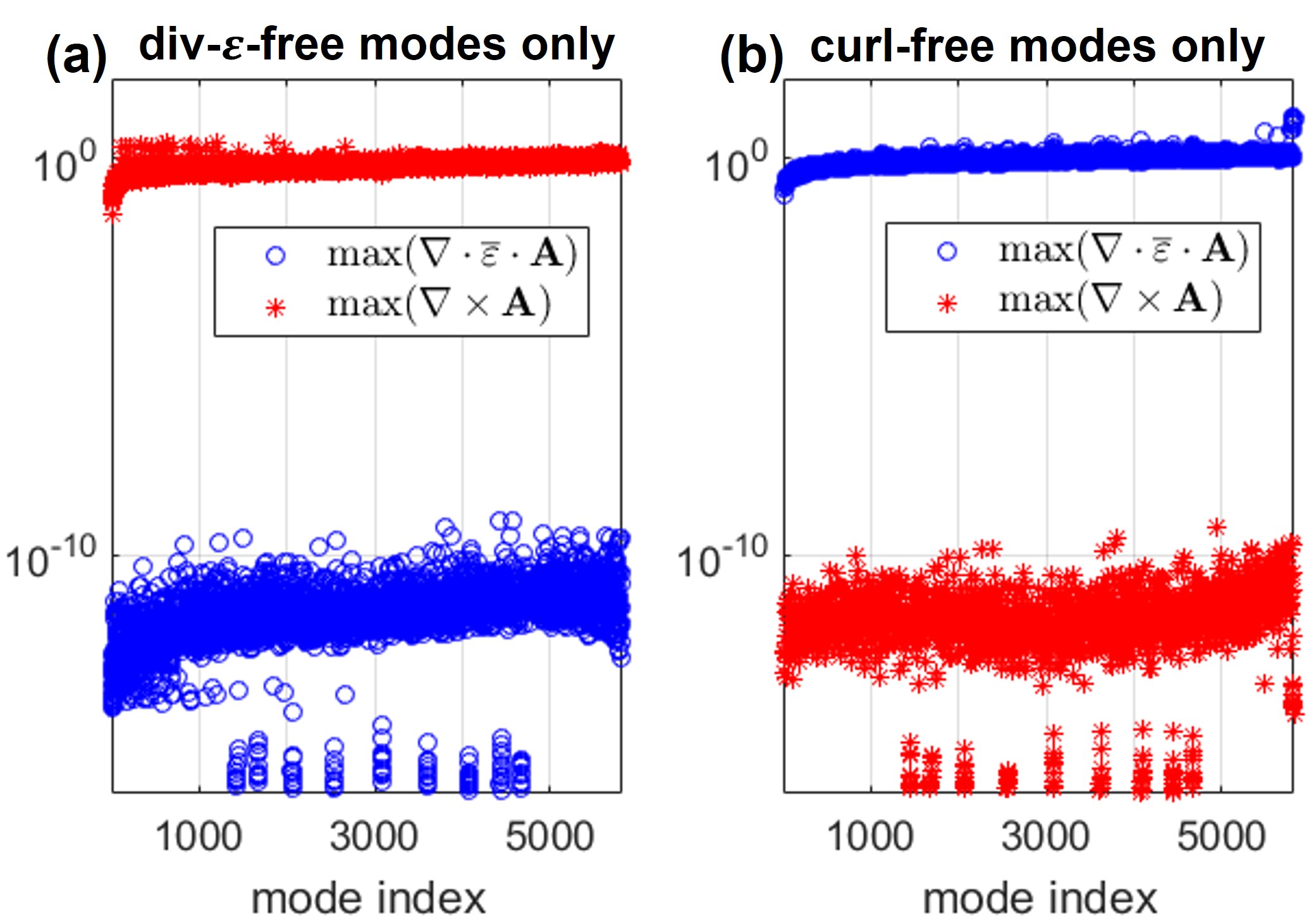}   
    \caption{Two set of modes found in the geometry described by Fig. \ref{fig:geometry}(b) can be clearly distinguished. 
    Totally there are 4049 div-$\epsilon$-free modes and 1773 curl-free modes.
    The description of the figure follows the caption of Fig. \ref{fig:numerical_proof_completeness_homo}(b) and (c). }
    \label{fig:numerical_proof_completeness_aniso_3}
\end{figure}

Mode degeneracy is observed after solving the generalized eigenvalue problem. We apply the analysis described in Appendix B to separate degenerate modes. 
The results are then displayed in Fig. \ref{fig:numerical_proof_completeness_aniso_3}. 
We find that the modes can be categorized into two distinct groups: 
\begin{enumerate}
    \item div-$\epsilon$-free (or generalized transverse) modes (Fig. \ref{fig:numerical_proof_completeness_aniso_3}(a)):
    $\nabla\cdot\epsilon\mathbf{A}=0$ for the entire grid; 
    \item Curl-free (or longitudinal) modes (Fig. \ref{fig:numerical_proof_completeness_aniso_3}(b)):
    $\nabla\times\mathbf{A}=0$ for the entire grid.\footnote{The curl-free modes can be eliminated by $\Phi=0$ gauge. This will be explained in future work.}
\end{enumerate}
It can be seen that each mode is either div-$\epsilon$-free, or curl-free. The div-$\epsilon$-free modes are consistent with the modes found by $\mathbf{E}$-$\mathbf{H}$ formulation since (\ref{wave_eq_A_evp}) returns to the wave equation of $\mathbf{E}$ field \cite{chew2014vector,raman2010photonic}. The curl-free modes have zero $\mathbf{E}$ and $\mathbf{H}$ field by the fact that $\mathbf{B}=\nabla\times\mathbf{A}=0$. 

In both numerical examples, the two classes of modes form a complete basis. 
Although the completeness of basis expanded by the eigenvectors in the discrete space from the numerical examples does not necessarily guarantee the completeness of the two sets of modes in the continuous Hilbert space, one could use convergence and existence of the numerical solution to a partial differential equation with finite difference approximation \cite{leveque2007finite} to imply completeness of these modes in the continuous Hilbert space. Thus the numerical proof could motivate further study in the continuous Hilbert space as future work.


\subsection{Numerical Validation of Orthogonality}\label{subsection:numerical_orthogonality}
\begin{figure}
    \centering
    \includegraphics[width=0.5\textwidth]{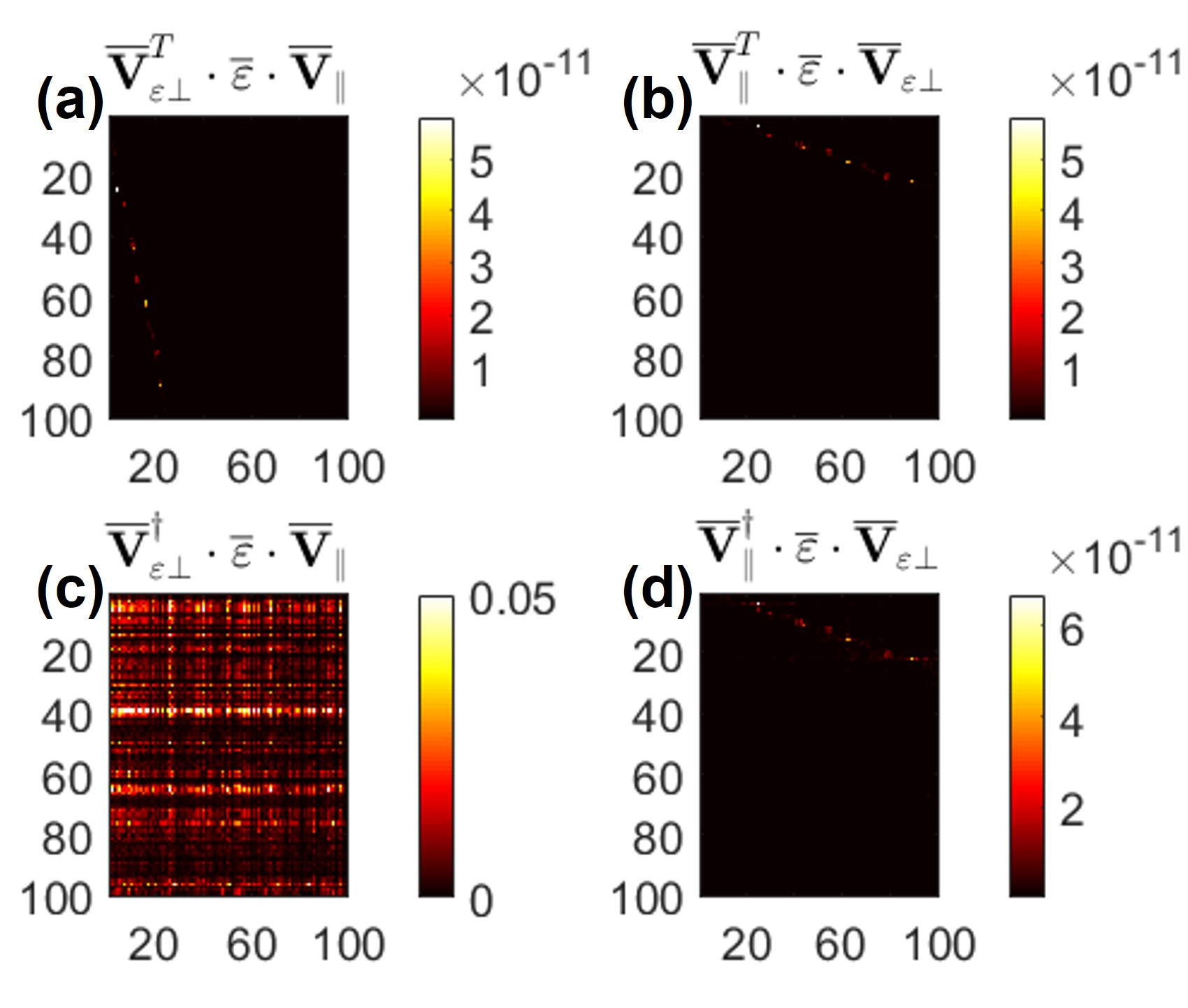}   
    \caption{Color map of four sets of inner products from the numerical example described by Fig. \ref{fig:geometry}(b). 
    (a) $\overline{\mathbf{V}}_{\epsilon\perp}^T \cdot \overline{\bm{\epsilon}} \cdot \overline{\mathbf{V}}_\parallel$; 
    (b) $\overline{\mathbf{V}}_\parallel^T \cdot \overline{\bm{\epsilon}} \cdot \overline{\mathbf{V}}_{\epsilon\perp}$;
    (c) $\overline{\mathbf{V}}_{\epsilon\perp}^\dagger \cdot \overline{\bm{\epsilon}} \cdot \overline{\mathbf{V}}_\parallel$;
    (d) $\overline{\mathbf{V}}_\parallel^\dagger \cdot \overline{\bm{\epsilon}} \cdot \overline{\mathbf{V}}_{\epsilon\perp}$.
    Here $\overline{\mathbf{V}}_{\epsilon\perp}$($\overline{\mathbf{V}}_\parallel$) is a matrix where each column vector is a div-$\epsilon$-free (curl-free) mode.
    Each pixel represents the magnitude of the corresponding matrix element.
    The indices are the indices of matrix elements.
    Note the different scales of the colorbars. The values of the resulting matrices are essentially 0 in (a), (b), and (d). }
    \label{fig:numerical_proof_orthogonality}
\end{figure}

We then proceed to validate the orthogonality conditions. For the numerical example described by Fig. \ref{fig:geometry}(b), we calculate the four sets of inner products between the div-$\epsilon$-free modes and curl-free modes defined in Section \ref{subsection:orthogonality_discrete} after separating the degenerate modes.
The colormaps of the resulting matrices are shown in Fig. \ref{fig:numerical_proof_orthogonality}, and the maximum magnitude of the matrix elements are shown in Table \ref{table_1}. 
For readability, only the first $100\times100$ elements of each matrix are displayed in Fig. \ref{fig:numerical_proof_orthogonality}, but these results are representative of the full matrices.
It can be seen from Fig. \ref{fig:numerical_proof_orthogonality} and Table \ref{table_1} that
\begin{align}
    \langle \mathbf{A}_{\epsilon\perp}, \epsilon \mathbf{A}_\parallel \rangle = 0 \label{orthogonality_1} 
\end{align}
\begin{align}
    \langle \mathbf{A}_\parallel, \epsilon \mathbf{A}_{\epsilon\perp} \rangle = 0 \label{orthogonality_2} 
\end{align}
\begin{align}
    \langle \mathbf{A}_{\epsilon\perp}^*, \epsilon \mathbf{A}_\parallel \rangle \neq 0 \label{orthogonality_3}
\end{align}
\begin{align}
    \langle \mathbf{A}_\parallel^*, \epsilon \mathbf{A}_{\epsilon\perp} \rangle = 0 \label{orthogonality_4} 
\end{align} 
which are consistent with the mathematical proofs in Section \ref{section:orthogonality}.
This result, supplemented by the numerical proof of completeness in the previous subsection, completes our demonstration of the generalized Helmholtz decomposition.

\begin{table}[h!]
\center
\caption{\label{table_1}
Orthogonality results from the second numerical example. The maximum magnitude of matrix elements in the corresponding matrices are shown in the table. Here $|\cdot|$ denotes element-wise magnitude.}
\begin{tabular}{|c | c |} 
 \hline
 max$(|\overline{\mathbf{V}}_{\epsilon\perp}^T \cdot \overline{\bm{\epsilon}} \cdot \overline{\mathbf{V}}_\parallel|)$ & 5.1694e-10 \\
 \hline 
 max$(|\overline{\mathbf{V}}_\parallel^T \cdot \overline{\bm{\epsilon}} \cdot \overline{\mathbf{V}}_{\epsilon\perp}|)$ & 5.1694e-10 \\
 \hline
 max$(|\overline{\mathbf{V}}_{\epsilon\perp}^\dagger \cdot \overline{\bm{\epsilon}} \cdot \overline{\mathbf{V}}_\parallel|)$ & 0.5779 \\
 \hline
 max$(|\overline{\mathbf{V}}_\parallel^\dagger \cdot \overline{\bm{\epsilon}} \cdot \overline{\mathbf{V}}_{\epsilon\perp}|)$ & 7.2430e-10 \\
 \hline
\end{tabular}
\end{table}


\section{Conclusion} \label{section:conclusion}
In this work, we combine mathematical proof and numerical results to demonstrate the generalization of Helmholtz decomposition to inhomogeneous media. 
The two families of fields in the generalized Helmholtz decomposition, div-$\epsilon$-free and curl-free fields, are connected to the two classes of solutions to the vector potential wave equation. 
The div-$\epsilon$-free and curl-free solutions form a complete set of the basis, as demonstrated in the numerical results and its discussion. They are also orthogonal to each other. 
The div-$\epsilon$-free field manifests charge-free condition, while the curl-free field is associated with charges.
In the literature, the curl-free fields are often eliminated by setting $\Phi=0$. However, when there are charges present, we need both families of solutions.
It also provides a theoretical background to the numerical quantization \cite{na2020quantum} based on the generalized Lorenz gauge. 
Future work will be devoted to the application of $\Phi=0$ gauge and rigorously determining suitable conditions for eliminating the curl-free modes in general inhomogeneous media while preserving full rank of the resulting system matrix for practical applications.

\appendices
\section{Discrete Vector Calculus on a Grid} \label{appendix:discrete}
Discrete vector calculus on a Yee lattice \cite{yee1966numerical,taflove2005computational} is the basis for the proof of orthogonality in discrete space. We give a brief overview of the discretization. The details can be found in \cite{chew1994electromagnetic}.

\begin{figure}
    \centering
    \includegraphics[width=0.25\textwidth]{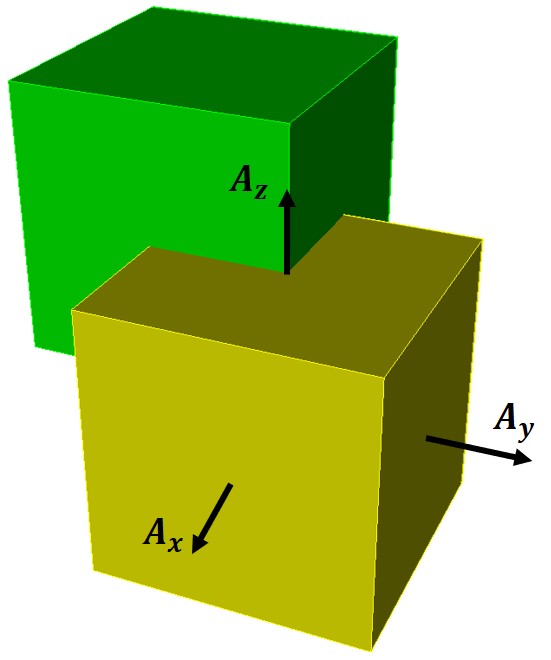}  
    \caption{Schematic of the grid. The primal and dual grids are shifted by half grid step.
    The solution vector $\mathbf{A}$ is placed on the face of the primal grid. } 
    \label{fig:grid}
\end{figure}

The staggered grid with both primal (solid line) and dual (dashed line) grids are shown in Fig. \ref{fig:grid}. 
The dual grid is shifted half a grid width in all three directions from the primal grid.
On each grid, we place a 3D vector on the face of the cube, and a scalar in the center of the cube. 

We index the grid point at the cube center of the primal grid by integers $(i,j,k)$. Below we explicitly formulate the discrete differential operations, where central difference is used. 

A discrete gradient $\mathbf{g}=\nabla f$ on the primal grid is given by
\begin{align}
    g_x^{i+0.5, j, k} &= \frac{f^{i+1,j,k}-f^{i,j,k}}{\Delta x} , \\
    g_y^{i, j+0.5, k} &= \frac{f^{i,j+1,k}-f^{i,j,k}}{\Delta y} , \\
    g_z^{i, j, k+0.5} &= \frac{f^{i,j,k+1}-f^{i,j,k}}{\Delta z} .
\end{align}

A discrete divergence $f = \nabla\cdot\mathbf{g}$ on the primal grid is given by
\begin{align}
    f^{i,j,k} &= \frac{g_x^{i+0.5, j, k} - g_x^{i-0.5, j, k}}{\Delta x} \nonumber \\
    &+ \frac{g_y^{i, j+0.5, k} - g_y^{i, j-0.5, k}}{\Delta y} \nonumber \\
    &+ \frac{g_z^{i, j, k+0.5} - g_z^{i, j, k-0.5}}{\Delta z} .
\end{align}

A discrete curl $\mathbf{h} = \nabla\times\mathbf{g}$ on a vector on the primal grid resulting in a vector on the dual grid is given by
\begin{align}
    h_x^{i,j+0.5,k+0.5} &= \frac{g_z^{i,j+1,k+0.5}-g_z^{i,j,k+0.5}}{\Delta y} \nonumber \\
    &- \frac{g_y^{i,j+0.5,k+1}-g_y^{i,j+0.5,k}}{\Delta z} , \\
    h_y^{i+0.5,j,k+0.5} &= \frac{g_x^{i+0.5,j,k+1}-g_x^{i+0.5,j,k}}{\Delta z} \nonumber \\
    &- \frac{g_z^{i+1,j,k+0.5}-g_z^{i,j,k+0.5}}{\Delta x} , \\
    h_z^{i+0.5,j+0.5,k} &= \frac{g_y^{i+1,j+0.5,k}-g_y^{i,j+0.5,k}}{\Delta x} \nonumber \nonumber \\
    &- \frac{g_x^{i+0.5,j+1,k}-g_x^{i+0.5,j,k}}{\Delta y} .
\end{align}

A discrete curl $\mathbf{g} = \nabla\times\mathbf{h}$ on a vector on the dual grid resulting in a vector on the primal grid is given by
\begin{align}
    g_x^{i+0.5,j,k} &= \frac{h_z^{i+0.5,j+0.5,k}-h_z^{i+0.5,j-0.5,k}}{\Delta y} \nonumber \\
    &- \frac{h_y^{i+0.5,j,k+0.5}-h_y^{i+0.5,j,k-0.5}}{\Delta z} , \\
    g_y^{i,j+0.5,k} &= \frac{h_x^{i,j+0.5,k+0.5}-h_x^{i,j+0.5,k-0.5}}{\Delta z} \nonumber \\
    &- \frac{h_z^{i+0.5,j+0.5,k}-h_z^{i-0.5,j+0.5,k}}{\Delta x} , \\
    g_z^{i,j,k+0.5} &= \frac{h_y^{i+0.5,j,k+0.5}-h_y^{i-0.5,j,k+0.5}}{\Delta x} \nonumber \nonumber \\
    &- \frac{h_x^{i,j+0.5,k+0.5}-h_x^{i,j-0.5,k+0.5}}{\Delta y} .
\end{align}

\section{Analysis of Degenerate Modes} \label{appendix:degeneracy}
Given a set of eigenmodes $\mathbf{A}_1, \mathbf{A}_2, \dots, \mathbf{A}_n$ with corresponding eigenvalues $\lambda_1, \lambda_2, \dots, \lambda_n$ obtained from the generalized eigenvalue problem, the space expanded by
\begin{align}
    \frac{1}{\lambda_i}\epsilon^{-1}\nabla\times\frac{1}{\mu}\nabla\times \mathbf{A}_i \label{M1_space}
\end{align}
has the same dimension as the number of independent div-$\epsilon$-free modes. This is because the curl-free modes are eliminated by the operation due to (\ref{null_A_parallel}) $\lambda_\parallel \nabla\times \mathbf{A}_\parallel=0$, and the div-$\epsilon$-free modes are recovered due to (\ref{evp_curl_curl}) $\nabla\times\frac{1}{\mu}\nabla\times \mathbf{A}_{\epsilon\perp} = \lambda_{\epsilon\perp} \epsilon \mathbf{A}_{\epsilon\perp}$.
Besides, the space expanded by
\begin{align}
    -\frac{1}{\lambda_i}\frac{1}{\epsilon_0^2\mu_0}\nabla\nabla\cdot\epsilon \mathbf{A}_i \label{M2_space}
\end{align}
has the same dimension as the number of independent curl-free modes. This is because the div-$\epsilon$-free modes are eliminated by the operation due to (\ref{null_A_perp}) $\lambda_{\epsilon\perp} \nabla\cdot\epsilon\mathbf{A}_{\epsilon\perp} = 0$, and the curl-free modes are recovered due to (\ref{evp_grad_div}) $- \frac{1}{\epsilon_0^2\mu_0} \epsilon\nabla\nabla\cdot\epsilon \mathbf{A}_\parallel = \lambda_\parallel \epsilon \mathbf{A}_\parallel$. 

\begin{figure}[h!]
    \centering
    \includegraphics[width=0.5\textwidth]{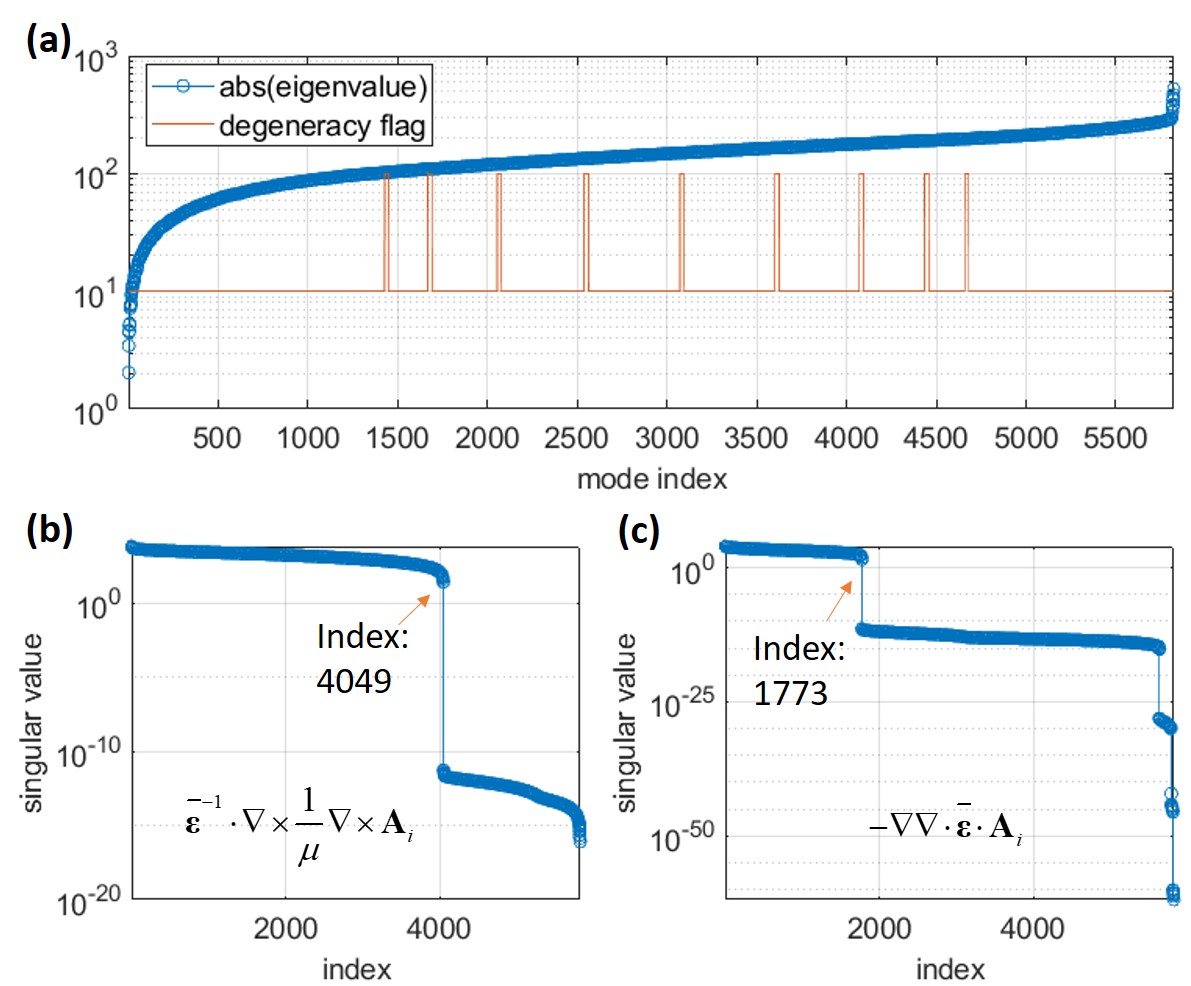}   
    \caption{The eigenvalues and the resulting solution space of the geometry described by Fig. \ref{fig:geometry}(b) are shown.
    (a) Eigenvalues are shown in blue curve; Orange upward spikes denote a cluster of degenerate modes.
    (b) The singular values of the matrix constructed by lining up $\overline{\bm{\epsilon}}^{-1}\cdot\nabla\times\frac{1}{\mu}\nabla\times\mathbf{A}_i$ are shown. It can be seen that there are 4049 nonzero singular values. 
    (c) The singular values of the matrix constructed by lining up $-\nabla\nabla\cdot\overline{\bm{\epsilon}}\cdot\mathbf{A}_i$ are shown. It can be seen that there are 1773 nonzero singular values. 
    The number of nonzero singular values indicate the dimension of the space spanned by a set of vectors.
    The total number of nonzero singular values from (b) and (c) is 5822, equal to the dimension of the problem. }
    \label{fig:numerical_proof_completeness_aniso_1}
\end{figure}

The dimension of the two spaces from the numerical example described by Fig. \ref{fig:geometry}(b) is shown in Fig. \ref{fig:numerical_proof_completeness_aniso_1}. 
The number of nonzero singular values indicate the dimension of the space spanned by a set of vectors.
It can be seen that the sum of the dimension of the space of (\ref{M1_space}) and (\ref{M2_space}) equals the dimension of the problem. Thus it proves that the div-$\epsilon$-free and curl-free modes as defined in the main text form a complete basis of the discrete space. 
Otherwise, if there exists an eigenmode that is nonzero after both (\ref{M1_space}) and (\ref{M2_space}) operations, it must contribute one ``extra dimension".  

\begin{figure}[h!]
    \centering
    \includegraphics[width=0.5\textwidth]{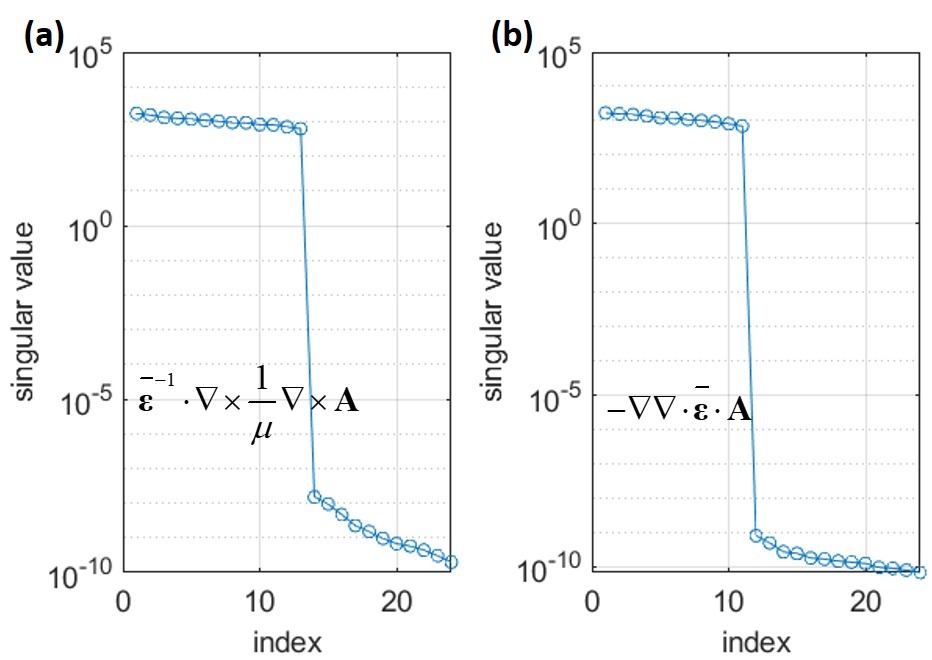}   
    \caption{A closer look at the first cluster of degenerate modes (index 1428 to 1451) in Fig. \ref{fig:numerical_proof_completeness_aniso_1}(a). Similar results are observed for the other clusters of degenerate modes.
    Singular values of (a) and (b) are obtained in the same way as Fig. \ref{fig:numerical_proof_completeness_aniso_1}(b) and (c). 
    The total number of nonzero singular values from (a) and (b) is 24, equal to the dimension of the degenerate subspace. }
    \label{fig:numerical_proof_completeness_aniso_2}
\end{figure}

Degenerate modes with the same eigenvalue can also be separated based on this analysis. 
Given $n$ degenerate modes $\mathbf{A}_1, \mathbf{A}_2, \dots, \mathbf{A}_n$ with the same eigenvalue, we first learn the number of independent div-$\epsilon$-free and curl-free modes by evaluating the dimension of the subspace expanded by (\ref{M1_space}) and (\ref{M2_space}), as shown in Fig. \ref{fig:numerical_proof_completeness_aniso_2}. 
The div-$\epsilon$-free modes are isolated by (\ref{M1_space}) because of (\ref{evp_curl_curl}), and the curl-free modes are isolated by (\ref{M2_space}) because of (\ref{evp_grad_div}).
Gram-Schmidt orthogonalization is then performed within each subspace to extract the independent div-$\epsilon$-free and curl-free modes, respectively.

\bibliographystyle{IEEEtran}
\bibliography{bibfile}

\end{document}